\documentclass[preprint,aps,12pt,showpacs,nofootinbib,tightenlines]{revtex4-1}
\usepackage{amsmath}
\usepackage{amssymb}
\usepackage{epsfig}
\usepackage{graphicx}
\newcommand{\non}{\nonumber\\ }

\def\be{\begin{equation}}
\def\ee{\end{equation}}
\def\ba{\begin{eqnarray}}
\def\ea{\end{eqnarray}}

\begin{document}
\title{ $B_s \to \phi l^+ l^-$ decays in the topcolor-assisted technicolor model }
\author{Lin-Xia L\"u$^1$\footnote{E-mail: lvlinxia@sina.com},~Xing-qiang Yang$^1$,
and Zong-chang Wang$^1$
\\
{\small $^1$ \it Physics and electronic engineering college, Nanyang
Normal University,} \\
{\small \it Nanyang, Henan 473061, P.R. China}  }

\begin{abstract}
Using the updated form factors within the light-cone QCD sum rule approach, we calculate the new physics contributions
to rare semileptonic $\bar{B}_s \to \phi\mu^+\mu^-, \phi \tau^+\tau^-$ decays from the new particles appearing
in the topcolor-assisted technicolor (TC2) model. In our evaluations, we find that: (i) the branching ratio,
normalized forward-backward asymmetry and lepton polarization asymmetries show highly sensitivity to charged
top-pions contributions and little sensitivity to $Z'$ contributions. The TC2 enhancements to the branching ratios
of these decays can reach a factor of $\sim 2$; (ii) the NP enhancement to the forward-backward asymmetry of the decay
$B_s\to \phi \mu^+\mu^-$ is in the range $-13\%$ to $3\%$, but $-9\%$ to $-6\%$ for decay $B_s\to \phi \tau^+\tau^-$
compared to the SM predictions; (iii) the TC2 model provide an enhancement of about $12\%$ to the longitudinal
polarization asymmetry $P_L$ for decay $B_s\to \phi \mu^+\mu^-$, but a decrease of about $10\%$ to the transverse
polarization asymmetry $P_T$ for the decay $B_s\to \phi \tau^+\tau^-$.

\end{abstract}


\maketitle

\section{Introduction}

High energy physics experiments are designed to resolve the yet-unanswered questions in the Standard Model~(SM) through searches of new physics~(NP) using  two approaches:
high energy or high luminosity approach. The first approach is to use high energy collider to
produce and discover new particles directly. The second one is to measure
flavor physics observables at for example B factory experiments and search for
the signal or evidence of a deviation from the SM prediction.
A natural place is to investigate the flavor-changing neutral current~(FCNC) processes in B meson rare decays.
In the SM, the rare B decays are all induced by the so-called box and/or penguin diagrams. Since these rare decay modes are
highly suppressed in the SM, they may serve as a good hunting ground for testing the SM and probing possible NP effects.

At the quark level, the decays $B_s \to \phi l^+ l^-$ proceed via FCNC transition $b \to s l^+ l^-$. The decays
$B_s \to \phi l^+ l^-$ will be one of the most important rare decays to be studied at the LHCb experiment and
other B physics experiments. The theoretically predicted branching ratio has the value of $\sim1.65\times10^{-6}$
for the $B_s \to \phi\mu^+\mu^-$ mode~\cite{Geng2003-sm}. The experimental observation was first done by
CDF collaboration~\cite{CDFPhill1} and the updated branching fraction is~\cite{CDFPhill2}
\begin{equation} \label{eq:Heff}
{\cal B}(\bar{B}_s \to \phi\mu^+\mu^-)=[1.47\pm0.24({\rm stat.}) \pm 0.46({\rm syst.})]\times10^{-6}.
\end{equation}
which is consistent with the SM prediction. However, the experimental errors are still large and this
decay allow for sizable NP contributions. We will evaluate the effects of possible
NP in these decays in the topcolor-assisted model.

The decays $B_s \to \phi l^+ l^-$ have been studied by employing the low-energy effective Hamiltonian and
nonperturbative approaches to compute the decay form factors in the framework of the SM~\cite{Geng2003-sm,Deandrea}.
Many studies about possible new physics contributions to these decays induced by loop diagrams involving various
new particles have been published, for example, in the two Higgs doublet model~\cite{Erkol}, the
universal extra dimension scenario~\cite{Mohanta}, the family non-universal $Z'$ model~\cite{qinchang} and other
new physics scenarios~\cite{Bsphill}.

The topcolor-assisted~(TC2)~\cite{Hill} model is one of the important candidates
for a mechanism of natural electroweak symmetry breaking. In the TC2 model,
the non-universal gauge boson~$Z'$, top-pions~$\pi_t^{0,\pm}$, top-Higgs boson~$h_t^0$
and other bound states may provide potentially large loop effects on low energy observables.
In this paper, we will investigate the new contributions from the new particles predicted by the
TC2 model to the branching ratios, the forward-backward asymmetry, and double lepton
polarization of the decays $B_s \to \phi l^+ l^-$.

The paper is arranged as follows. In Section~II, we give a brief review of the topcolor-assisted
technicolor model. In Section~III, we present the theoretical framework for $B_s\to \phi l^+ l^-$
decays within the SM and the TC2 model, then give the definitions and the derivations of the form
factors in the decays $B_s \to \phi l^+ l^-$ using the updated form factors within the light-cone
QCD sum rule. In Section~IV, we introduce the basic formula for experimental observables,
including dilepton invariant mass spectrum, forward-backward asymmetry~(FBA), and lepton
polarization. In Section~V, we present our numerical results for these decays in the SM and the TC2 model.
The conclusions are presented in the final section.

\section{Outline of the TC2 model }\label{TC2 model}

To completely avoid the problems arising from the elementary Higgs field in the SM, various kinds of dynamical
electroweak symmetry breaking (EWSB) models have been proposed, among which the topcolor scenario is attractive
because it can explain the large top quark mass and provide a possible EWSB mechanism~\cite{Hill}. Almost all
of these kinds of models propose that the underlying interactions, topcolor interactions, should be flavor
nonuniversal. These non-universal interactions in the mass eigenstate basis generate tree-level flavor
changing (FC) couplings and result in a rich phenomenology.

A key feature of the TC2 model is the presence of top-pions($\pi_t^{0,\pm}$), the non-universal
gauge boson ($Z'$) and the top-Higgs ($h_t^0$). These new particles
treat the third-generation fermions differently from those in the first and second generations
and thus can lead to tree-level flavor changing (FC) couplings.
When one writes the non-universal interactions in the quark mass
eigen-basis, the FC couplings of top-pions to quarks can be written as
\cite{tc201,tc204}:
\begin{eqnarray}
&&\frac{m_{t}^*}{\sqrt{2}F_{\pi}}\frac{\sqrt{\nu_{w}^{2}-
F_{\pi}^{2}}}{\nu_{w}}\left[iK_{UR}^{tc}K_{UL}^{tt^{*}}\bar{t}_{L}c_{R}\pi_{t}^{0}
+\sqrt{2}K_{UR}^{tc^{*}}K_{DL}^{bb}\bar{c}_{R}b_{L}\pi_{t}^{+}
+\sqrt{2}K_{UR}^{tc}K_{DL}^{bb^{*}}\bar{b}_{L}c_{R}\pi_{t}^{-}\right.\nonumber \\
&&\hspace{15mm}\left.+\sqrt{2}K_{UR}^{tc^{*}}K_{DL}^{ss}\bar{t}_{R}s_{L}\pi_{t}^{+}
+\sqrt{2}K_{UR}^{tc}K_{DL}^{ss^{*}}\bar{s}_{L}t_{R}\pi_{t}^{-}\right].
\end{eqnarray}
Here $\nu_{w}=\nu/\sqrt{2}=174\rm~GeV$, $F_{\pi}\approx 50 {\rm GeV}$ is the top-pion decay constant,
$K_{UL(R)}$ and $K_{DL(R)}$ are rotation matrices and satisfy the equations $K_{UL}^{+} M_{U}K_{UR}=M_{U}^{dia}$
and $K_{DL}^{+}M_{D}K_{DR}=M_{D}^{dia}$, in which $M_{U}$ and $M_{D}$ are up-quark and down-quark mass
matrices, respectively. The values of the coupling parameters can be taken as~\cite{tc204}:
\begin{equation}
K_{UL}^{tt}\approx K_{DL}^{bb} \approx K_{DL}^{ss}\approx1,
\hspace{10mm} K_{UR}^{tc}\leq\sqrt{2\varepsilon-\varepsilon^{2}} .
\end{equation}
In our numerical analysis, we will take
$K_{UR}^{tc}=\sqrt{2\varepsilon- \varepsilon^{2}}$.
The predicted top-Higgs $h_t^0$ is a
$t\bar t$ bound analogous to the $\sigma $ particle in low
energy QCD, and its Feynman rules are similar to the SM Higgs boson.

The Flavor diagonal (FD) couplings of top-pions to fermions
take the form~\cite{Hill,tc203}:
\begin{eqnarray}
&&\frac{m_t^*}{\sqrt{2}F_{\pi}} \frac{\sqrt{\nu_{w}^{2}-
F_{\pi}^{2}}}{\nu_{w}}\left[i\bar{t}\gamma^{5}t\pi_{t}^{0}+\sqrt{2}\bar{t}_{R}b_{L}
\pi_{t}^{+}+\sqrt{2}\bar{b}_{L}t_{R}\pi_{t}^{-}\right] \nonumber \\
&&+\frac{m_{b}^{*}}{\sqrt{2}F_{\pi}}\left[i\bar{b}\gamma^{5}b\pi_{t}^{0}+\sqrt{2}
\bar{t}_{L}b_{R}\pi_{t}^{+}+\sqrt{2}\bar{b}_{R}t_{L}\pi_{t}^{-}\right]
+\frac{m_{l}}{\nu}\bar{l}\gamma^{5}l\pi^{0}_{t}.
\end{eqnarray}
The TC2 model postulates that topcolor interactions mainly couple to the third generation
fermions, and give rise to the main part of the quark mass $m_t^*=m_{t}(1-\varepsilon)$,
while the masses of the ordinary
fermions are induced by ETC (extended technicolor) interactions with $m_b^*=m_b-0.1\varepsilon m_t$.

The FC couplings of the non-universal gauge boson $Z'$ to fermions, which may
provide significant contributions to some FCNC processes, can be written as
\cite{tc210}:
\begin{eqnarray}
  {\cal L}^{FC}_{Z'}=-\frac{g_1}{2}\cot{\theta'} Z'^{\mu}\left\{\frac{1}{3}D_{L}^{bb} D_{L}^{bs*}
  \bar{s}_L\gamma_{\mu} b_L -
\frac{2}{3}D_{R}^{bb}D_{R}^{bs*}\bar{s}_R \gamma_{\mu} b_R  +{\rm
h.c.} \right\} \label{bsz},
\end{eqnarray}
Here $g_{1}$ is the ordinary
hypercharge gauge coupling constant,
$D_L,D_R$ are matrices which rotate the weak eigen-basis to the mass eigen-basis
for the down-type left and right hand quarks. The FD
couplings of $Z'$ to fermions can be written as~\cite{Hill,tc201,tc203}:
\begin{eqnarray}
{\cal L}^{FD}_{Z'}&=&-\sqrt{4\pi K_{1}}\left\{
Z'_{\mu}\left[\frac{1}{2}\bar{\tau}_{L}\gamma^{\mu}\tau_{L}
-\bar{\tau}_{R}\gamma^{\mu}\tau_{R}+\frac{1}{6}\bar{t}_{L}\gamma^{\mu}t_{L}
+\frac{1}{6}\bar{b}_{L}\gamma^{\mu}b_{L}+\frac{2}{3}\bar{t}_{R}\gamma^{\mu}t_{R}
\right.\right.\nonumber\\
&-&\left.\left.\frac{1}{3}\bar{b}_{R}\gamma^{\mu}b_{R}\right]-\tan^{2}
\theta'Z'_{\mu}\left[\frac{1}{6}\bar{s}_{L}\gamma^{\mu}s_{L}
-\frac{1}{3}\bar{s}_{R}\gamma^{\mu}s_{R}-\frac{1}{2}
\bar{\mu}_{L}\gamma^{\mu}\mu_{L}-\bar{\mu}_{R}\gamma^{\mu}\mu_{R}
\right.\right.\nonumber\\
&-&\left.\left.\frac{1}{2}\bar{e}_{L}\gamma^{\mu}e_{L}-\bar{e}_{R}\gamma^{\mu}e_{R}\right]\right\}.
\label{zll}\end{eqnarray}
Here $\theta'$ is the mixing angle, and $K_{1}$ is the coupling constant with $\tan
\theta'=\frac{g_{1}}{\sqrt{4\pi K_{1}}}$.

\section{Effective Hamiltonian and form factors}\label{Hamiltonian}

In the TC2 model, after neglecting the doubly Cabibbo-suppressed contributions, the effective hamiltonian
for the transition $b \rightarrow s l^+ l^-$ has the following structure~\cite{dai,wenjunli}:
\be \label{Heff} {\cal H} = - \frac{4 G_F}{\sqrt{2}} V^*_{ts}
V_{tb} \sum \limits_{i=1}^{10} [C_i(\mu) {\cal O}_i(\mu)+
C_{Q_i}(\mu) Q_i(\mu) ] \ee
where $V^*_{ts} V_{tb}$ is the CKM
factor, and $G_F$ is the Fermi coupling constant.
$C_i$ and $C_{Q_i}$ are the Wilson coefficients at the renormalization point
$\mu=m_W$, ${\cal O}_i$'s ($i=1,\cdots,10$) are the operators in the SM and the explicit expressions
can be found in Ref.~\cite{buras52}, and $Q_i$'s come from the diagrams exchanging the neutral particles in TC2
and are~\cite{dai,wenjunli}
\begin{align}
Q_1 &
=\frac{e^2}{16\pi^2}(\bar{s}^{\alpha}_Lb^{\alpha}_R)(\bar{l}
l)\, , & Q_2& =
\frac{e^2}{16\pi^2}(\bar{s}^{\alpha}_Lb^{\alpha}_R)(\bar{l}\gamma_5
l)\, , \non
Q_{3}&=\frac{g_s^2}{16\pi^2}(\bar{s}^{\alpha}_Lb^{\alpha}_R)\left(\sum_q\bar{q}^{\beta}
_{L}q^{\beta}_{R}\right)\, ,&
Q_4&=\frac{g_s^2}{16\pi^2}(\bar{s}^{\alpha}_Lb^{\alpha}_R)\left(\sum_q\bar{q}^{\beta}
_Rq^{\beta}_L\right)\, , \non
Q_{5}&=\frac{g_s^2}{16\pi^2}(\bar{s}^{\alpha}_Lb^{\beta}_R)\left(\sum_q\bar{q}^{\beta}
_{L}q^{\alpha}_{R}\right)\, ,&
Q_6&=\frac{g_s^2}{16\pi^2}(\bar{s}^{\alpha}_Lb^{\beta}_R)\left(\sum_q\bar{q}^{\beta}
_Rq^{\alpha}_L\right)\, , \non
Q_7&=\frac{g_s^2}{16\pi^2}(\bar{s}^{\alpha}_L\sigma^{\mu\nu}b^{\alpha}_R)
\left(\sum_q\bar{q}^{\beta}_L\sigma_{\mu\nu}q^{\beta}_R\right)\, ,&
Q_8&=\frac{g_s^2}{16\pi^2}(\bar{s}^{\alpha}_L\sigma^{\mu\nu}b^{\alpha}_R)
\left(\sum_q\bar{q}^{\beta}_R\sigma_{\mu\nu}q^{\beta}_L\right)\, ,
\non
Q_9&=\frac{g_s^2}{16\pi^2}(\bar{s}^{\alpha}_L\sigma^{\mu\nu}b^{\beta}_R)
\left(\sum_q\bar{q}^{\beta}_L\sigma_{\mu\nu}q^{\alpha}_R\right)\, ,&
Q_{10}&=\frac{g_s^2}{16\pi^2}(\bar{s}^{\alpha}_L\sigma^{\mu\nu}b^{\beta}_R)
\left(\sum_q\bar{q}^{\beta}_R\sigma_{\mu\nu}q^{\alpha}_L\right)\, .
\label{oper}
\end{align}
where $\alpha$ and $\beta$ denote color indices. The subscripts
$L$ and $R$ refer to left- and right- handed components of the
fermion fields. $e$ and $g_s$ are the electromagnetic and strong
coupling constants respectively.

In terms of the above effective Hamiltonian~(\ref{Heff}),
the decay amplitude of $b \rightarrow s l^+ l^-$ can be written  as~\cite{wenjunli}:
\ba {\cal M}&=&\frac{G_F\alpha_{em}}{2\sqrt{2}\pi}V_{tb}V_{ts}^*
\Bigg \{-2\widetilde C_{7}^{eff}\hat{m}_b\bar{s}i\sigma_{\mu\nu}
\frac{\hat{q}^\nu}{\hat{s}}(1+\gamma_5)b\bar{l}\gamma^\mu l +
\widetilde C_{9}^{eff}\bar{s}\gamma_\mu(1-\gamma_5)
b\bar{l}\gamma^\mu l  \non &&  + \widetilde
C_{10}^{eff}\bar{s}\gamma_\mu(1-\gamma_5)
b\bar{l}\gamma^\mu\gamma_5l +
C_{Q_1}\bar{s}(1+\gamma_5)b\bar{l}l
+C_{Q_2}\bar{s}(1+\gamma_5)b\bar{l}\gamma_5l \Bigg \}.
\label{matrix}
\ea
In the SM, the effective Wilson coefficients
which enter the decay distributions are written as~\cite{buras52}
\be\label{eq:effWC}
 C_9^{\rm eff}(\hat{s}) = C_9 + Y(\hat{s})\,,
\ee
in which $Y(\hat{s})$ stands for the matrix element of four-quark operators and given by
\ba
Y(\hat{s})&=&h(z,\hat{s})\big(3C_1+C_2+3C_3+C_4+3C_5+C_6\big)-
\frac{1}{2}h(1,\hat{s})\big(4C_3+4C_4+3C_5+C_6\big)\,\nonumber\\
& &-\frac{1}{2}h(0,\hat{s})\big(C_3+3C_4\big)+\frac{2}{9}\big(3C_3+C_4+3C_5+C_6\big)\,.
\ea
Here the long-distance contributions from the resonant states have been neglected because
they could be excluded by experimental analysis~\cite{CDFPhill1}.
The detailed discussion of the resonance effects can be found in Ref.~\cite{resEff}.

In the TC2 model, After the breaking of the extended gauge group to their
diagonal subgroups, the non-universal massive gauge boson~$Z'$ is produced.
It generally couples to the third-generation fermions and have large tree-level
flavor changing couplings. The non-universal gauge boson~$Z'$ can give a
correction to the function $C_0(x)$ of the SM~\cite{smf}. For $l=e, \mu$, the $C_{01}^{TC2}(x)$
is~\cite{chongxingyue99}
\ba
C_{01}^{TC2}(y_t)&=&\frac{-tan^2\theta'
M_Z^2}{M_{Z'}^2}\left[K_{ab}(y_t)+K_c(y_t)+K_d(y_t)\right]\label{ytcz},
\ea
with $y_t=m_t^{*2}/M_W^2$. For the decay process $B_s \to \phi \tau^+\tau^- $, the factor $-tan^2\theta'$
should be replaced by 1.
For the convenience of the reader, we present the functions $K_{ab}(y_t)$, $K_c(y_t)$ and
$K_d(y_t)$ in the Appendix~\ref{app:aa}.

The charged top-pions $\pi_t^{\pm}$ can give contributions to the corresponding
SM functions $C_0(x)$, $D_0(x)$, $E_0(x)$ and $E'_0(x)$. The explicit expressions
of these functions are~\cite{xiao2001}:
\ba
C_{02}^{TC2}(z_t)&=&\frac{m_{\pi}^2}{4\sqrt{2}G_F M_W^2
F_{\pi}^2}\left[-\frac{z_t^2}{8(1-z_t)}-\frac{z_t^2}{8(1-z_t)^2}{\rm log}[z_t]\right],\\
D_{0}^{TC2}(z_t)&=&\frac{1}{4\sqrt{2}G_FF_{\pi}^2}\left[\frac{47-79z_t+38z_t^2}{108(1-z_t)^3}+\frac{3-6z_t^2+4z_t^3}
{18(1-z_t)^4}{\rm log}[z_t]\right],\\
E_{0}^{TC2}(z_t)&=&\frac{1}{4\sqrt{2}G_FF_{\pi}^2}\left[\frac{7-29z_t+16z_t^2}{36(1-z_t)^3}-\frac{3z_t^2-2z_t^3}{6(1-z_t)^4}
{\rm log}[z_t]\right],\\
E_{0}^{'TC2}(z_t)&=&\frac{1}{8\sqrt{2}G_FF_{\pi}^2}\left[-\frac{5-19z_t+20z_t^2}{6(1-z_t)^3}+\frac{z_t^2-2z_t^3}{(1-z_t)^4}
{\rm log}[z_t]\right].
\ea
Here $z_t={m^*_t}^2/m_{\pi_t^{\pm}}^2$.

The neutral top-pion $\pi_t^0$ and top-Higgs $h_t^0$ can also give contributions to the rare decays $B_s \to \phi l^+ l^-$
through the new operators given in Eq.~(\ref{oper})~\cite{chongxingyue99}. The corresponding Wilson
coefficients are written as:
\begin{equation}
C_{Q_1} = \frac{\sqrt{\nu_w^2-F_{\pi}^2}}{\nu_w}\left[
\frac{m_b^*m_l\nu} {2\sqrt{2}sin^2\theta_wF_{\pi}m_{\pi_t^0}^2
}C_0(x_t)+\frac{V_{ts}m_lm_t^*m_b^{*2}M_W^2 }{4\sqrt{2}\nu g_2^4
F_{\pi}^3m_{\pi_t^0}^2 }C(x_s)\right]. \label{rs}
\end{equation}
Here $x_s={m^*_t}^2/m_{\pi_t^0}^2$, $g_2$ is the $SU(2)$ coupling constant, and
$C_0(x_t)$ is the Inami-Lim function in the SM~\cite{smf}. The expression of $C_{Q_2}$ is same as that
of $C_{Q_1}$ except for the masses of the scalar particles.

Exclusive decays are described in terms of matrix
elements of the quark operators in Eq.~(\ref{matrix}) over meson
states, which are described by several independent form factors.
For $B_s \to \phi l^+ l^-$, the related transition matrix elements are defined
as $(q=p-k)$~\cite{Ball:2004rg}
\ba \langle \phi(k) | (V-A)_\mu | B(p)\rangle & = & -i
\epsilon^*_\mu (m_{B_s}+m_{\phi}) A_1(s) + i (p+k)_\mu (\epsilon^*
p)\,
\frac{A_2(s)}{m_{B_s}+m_{\phi}}\nonumber\\
\lefteqn{+ i q_\mu (\epsilon^* p) \,\frac{2m_{\phi}}{s}\,
\left(A_3(s)-A_0(s)\right) +
\epsilon_{\mu\nu\rho\sigma}\epsilon^{*\nu} p^\rho k^\sigma\,
\frac{2V(s)}{m_{B_s}+m_{\phi}}\,.}\hspace*{2cm}\label{eq:ff3} \ea
with $A_3(s) = \frac{m_{B_s}+m_{\phi}}{2m_{\phi}}\, A_1(s) -
\frac{m_{B_s}-m_{\phi}}{2m_{\phi}}\, A_2(s)$ and $A_0(0) =  A_3(0)$,
\ba \langle {\phi(k)} | \bar s \sigma_{\mu\nu} q^\nu (1+\gamma_5) b
| B(p)\rangle & = & i\epsilon_{\mu\nu\rho\sigma} \epsilon^{*\nu}
p^\rho k^\sigma \, 2 T_1(s)\nonumber\\
& & {} + T_2(s) \left\{ \epsilon^*_\mu
  (m_{B_s}^2-m_{\phi}^2) - (\epsilon^* k) \,(p+k)_\mu \right\}\nonumber\\
& & {} + T_3(s) (\epsilon^* p) \left\{ q_\mu -
\frac{s}{m_{B_s}^2-m_{\phi}^2}\, (p+k)_\mu \right\}.\label{eq:T} \ea
with $T_1(0) = T_2(0)$. $\epsilon_\mu$ is the polarization vector of the $\phi$ meson. The physical range
in $s=q^2$ extends from $s_{\rm min} = 0$ to $s_{\rm max} =(m_{B_s}-m_{\phi})^2$.

\begin{table}[t]
\begin{center}
\caption{\label{FFfit} Form factors for $B_s\to\phi$ transition within the light-cone QCD sum rule.}
\vspace{0.3cm}
\begin{tabular}{crrrrr}\hline\hline
                         &$F(0)$   &$r_1$   &$m_R^2$   &$r_2$     &$m^2_{\rm fit}$ \\ \hline
$V^{B_s\to\phi}$         &$0.434$  &$1.484$ &$5.32^2$  &$-1.049$  &$39.52$         \\\hline
$A_0^{B_s\to\phi}$       &$0.474$  &$3.310$ &$5.28^2$  &$-2.835$  &$31.57$         \\\hline
$A_1^{B_s\to\phi}$       &$0.311$  &---     &---       &$0.308$   &$36.54$         \\\hline
$A_2^{B_s\to\phi}$       &$0.234$  &$-0.054$&---       &$0.288$   &$48.94$         \\\hline
$T_1^{B_s\to\phi}$       &$0.349$  &$1.303$ &$5.32^2$  &$-0.954$  &$38.28$         \\\hline
$T_2^{B_s\to\phi}$       &$0.349$  &---     &---       &$0.349$   &$37.21$         \\\hline
$\tilde{T}_3^{B_s\to \phi}$&$0.349$  &$0.027$ &---       &$0.321$   &$45.56$        \\
\hline\hline
\end{tabular}
\end{center}
\end{table}
Form factors for $B_s\to\phi$ transition have been updated recently in the light-cone QCD sum rule
approach~\cite{Ball:2004rg}. For the $q^2$ dependence of the form factors, they have been
parameterized by a simple formulae with two or three parameters. The form factors
$V$, $A_0$ and $T_1$ are parameterized by
\begin{eqnarray}
F(s)=\frac{r_1}{1-s/m^2_{R}}+\frac{r_2}{1-s/m^2_{\rm fit}},
\label{r12mRfit}
\end{eqnarray}
For the form factors $A_2$ and $\tilde{T}_3$, it can be expanded to the second order around the pole, giving
\begin{eqnarray}
F(s)=\frac{r_1}{1-s/m^2}+\frac{r_2}{(1-s/m)^2}\,, \label{r12mfit}
\end{eqnarray}
where $m=m_{\rm fit}$ for $A_2$ and $\tilde{T}_3$. The fit formula for $A_1$ and $T_2$ is
\begin{eqnarray}
F(s)=\frac{r_2}{1-s/m^2_{\rm fit}}.\label{r2mfit}
\end{eqnarray}
The form factor $T_3$ can be obtained by $T_3(s)=\frac{m_{B_s}^2-m_{\phi}^2}{s}\big[\tilde{T}_3(s)-T_2(s)\big]$. All of the
form factors are collected Table~\ref{FFfit}.

\section{Basic Formula for Observables}
In this section, we give formula for experimental observables
including dilepton invariant mass spectrum, forward-backward asymmetry~(FBA), and lepton
polarization.

From Eqs.~(\ref{matrix}-\ref{eq:T}), the decay matrix element of
$B_s \to \phi l^+ l^-$ can be written in the form
\be
{\cal M}
=-\frac{G_F\alpha_{em}}{2\sqrt{2}\pi}V_{tb}V^*_{ts}m_{B_s}\left[{\cal
T}^1_\mu(\overline{l}\gamma^{\mu}l) +{\cal T}^2_\mu
(\overline{l} \gamma^{\mu}\gamma_5 l)+{\cal
S}(\overline{l} l)
   \right\}
\ee
with
\ba {\cal T}^1_\mu &=&
A(\hat{s})\epsilon_{\mu\rho\alpha\beta}\epsilon^{*\rho}
\hat{p}^\alpha \hat{k}^\beta
-iB(\hat{s})\epsilon^{*}_{\mu}+iC(\hat{s})(\epsilon^{*} \cdot
\hat{p})(\hat{p}+\hat{k})_{\mu},\\
{\cal T}^2_\mu
&=&E(\hat{s})\epsilon_{\mu\rho\alpha\beta}\epsilon^{*\rho}
\hat{p}^\alpha \hat{k}^\beta-iF(\hat{s})\epsilon^*_\mu
+iG(\hat{s})(\epsilon^* \cdot \hat{p})(\hat{p}+\hat{k})_\mu
+iH(\hat{s})(\epsilon^{*} \cdot \hat{p})\hat{q}_\mu, \\
{\cal
S}&=&i2\hat{m}_{\phi}(\epsilon^* \cdot \hat{p}){\cal
S}_2(\hat{s}) \ea
where $\hat{m}=\frac{m}{m_{B_s}}$, $\hat{p}=\frac{p}{m_{B_s}}$, and the auxiliary functions
are then given by:
\ba
A(\hat{s})&=&\frac{2}{1+\hat{m}_{\phi}} \widetilde
{C}_9^{eff}(\hat{s})V(\hat{s})
+\frac{4\hat{m}_b}{\hat{s}}\widetilde C_7^{eff} T_1(\hat{s}),\label{ahs}\\
B(\hat{s})&=&(1+\hat{m}_{\phi})\widetilde
{C}_9^{eff}(\hat{s})A_1(\hat{s})
+ \frac{2\hat{m}_b}{\hat{s}}(1-\hat{m}^2_{\phi})\widetilde C_7^{eff}T_2(\hat{s}),
\label{bhs}\\
C(\hat{s})&=&\frac{1}{1+\hat{m}_{\phi}}\widetilde
{C}_9^{eff}(\hat{s})A_2(\hat{s})
+\frac{2\hat{m}_b}{1-\hat{m}^2_{\phi}}\widetilde
C_7^{eff}\left(T_3(\hat{s})+
\frac{1-\hat{m}^2_{\phi}}{\hat{s}}T_2(\hat{s})\right),\label{chs}\\
E(\hat{s})&=&\frac{2}{1+\hat{m}_{\phi}}
\widetilde C_{10}^{eff}V(\hat{s}),\label{ehs}\\
F(\hat{s})&=&(1+\hat{m}_{\phi})\widetilde C_{10}^{eff}A_1(\hat{s}),\label{fhs}\\
G(\hat{s})&=&\frac{1}{1+\hat{m}_{\phi}}\widetilde C_{10}^{eff}A_2(\hat{s}),\label{ghs}\\
H(\hat{s})&=&\frac{2\hat{m}_{\phi}}{\hat{s}}\widetilde C_{10}^{eff}
\left(A_3(\hat{s})-A_0(\hat{s})\right)
+\frac{\hat{m}_{\phi}}{\hat{m}_l(\hat{m}_b+\hat{m}_s)}C_{Q_2}A_0(\hat{s}),\label{hhs}\\
{\cal
S}_2(\hat{s})&=&-\frac{1}{(\hat{m}_b+\hat{m}_s)}A_0(\hat{s})C_{Q_1}.\label{s2s}
\ea

The contributions of $Z'$ and charged top-pions are translated through
the RGE step into modifications of the effective Wilson
coefficients $\widetilde {C}_7^{eff}$, $\widetilde {C}_9^{eff}$
and $\widetilde {C}_{10}^{eff}$, while the contributions of
neutral top-pion and top-Higgs are incorporated in the terms of $H(\hat{s})$ and ${\cal
S}_2(\hat{s})$.

The two kinematic variables $\hat{s}$ and $\hat{u}$ are chosen to be
\ba \label{su}
\hat{s} &=&\hat{q}^2=(\hat{p}_++\hat{p}_-)^2, \\
\hat{u}&=&(\hat{p}-\hat{p}_-)^2-(\hat{p}-\hat{p}_+)^2, \ea
which are bounded as
\ba
(2\hat{m}_l)^2\leq&\hat{s}&\leq(1-\hat{m}_{\phi})^2,\label{slimit}\\
-\hat{u}(\hat{s})\leq&\hat{u}&\leq\hat{u}(\hat{s}),
\label{usbound}\ea
with $\hat{m}_l=m_{l}/m_B$. Here the variable $\hat{u}$ is
related to the angle $\theta$ between the momentum of the
$B$-meson and that of $l^+$ in the center of mass frame of the
dileptons $l^+l^-$ through the relation
$\hat{u}=-\hat{u}(\hat{s})\cos\theta$. $\hat{u}(\hat{s})$ can be
written as follows
\be
\hat{u}(\hat{s})=\sqrt{\lambda\big(1-4\frac{\hat{m}^2_{l}}{\hat{s}}\big)},\ee
with \ba \lambda&\equiv &\lambda(1,\hat{m}^2_{\phi},\hat{s})\non
&=&1+\hat{m}^4_{\phi}
+\hat{s}^2-2\hat{s}-2\hat{m}^2_{\phi}(1+\hat{s}). \ea

Keeping the lepton mass and integrating over $\hat{u}$ in the
kinematic region given in Eq.~(\ref{usbound}), we can obtain the
differential decay rates for the decays $B_s \to \phi l^+ l^-$:
\ba \frac{dBr}{d\hat{s}}&=&\tau_{B_s}
\frac{G^2_F\alpha_{em}^2m^5_{B_s}}{2^{10}\pi^5}
|V_{tb}V^*_{ts}|^2 \hat{u}(\hat{s})D^{\phi}, \label{ims}\\
D^{\phi}&=&
\frac{|A|^2}{3}\hat{s}\lambda(1+2\frac{\hat{m}^2_l}{\hat{s}})
+\frac{|E|^2}{3}\hat{s}\hat{u}(\hat{s})^2 +|{\cal
S}_2|^2(\hat{s}-4\hat{m}^2_l)\lambda \non
&&+\frac{1}{4\hat{m}^2_{\phi}}\left[|B|^2(\lambda-\frac{\hat{u}(\hat{s})^2}{3}
+8\hat{m}^2_{\phi}(\hat{s}+2\hat{m}^2_l))
+|F|^2(\lambda-\frac{\hat{u}(\hat{s})^2}{3}+8\hat{m}^2_{\phi}
(\hat{s}-4\hat{m}^2_l))\right]  \non
&&+\frac{\lambda}{4\hat{m}^2_{\phi}}\left[|C|^2(\lambda-\frac{\hat{u}(\hat{s})^2}{3})
+|G|^2\left(\lambda-\frac{\hat{u}(\hat{s})^2}{3}+4\hat{m}^2_l
(2+2\hat{m}^2_{\phi}-\hat{s})\right)\right]\non
&&-\frac{1}{2\hat{m}^2_{\phi}}\left[Re(BC^{*})(1-\hat{m}^2_{\phi}-\hat{s})
(\lambda-\frac{\hat{u}(\hat{s})^2}{3})\right.\non
&&\left.+Re(FG^{*})\left((1-\hat{m}^2_{\phi}-\hat{s})
(\lambda-\frac{\hat{u}(\hat{s})^2}{3})+4\hat{m}^2_l\lambda\right)\right]
\non
&&-2\frac{\hat{m}^2_l}{\hat{m}^2_{\phi}}\lambda\left[Re(FH^{*})
-Re(GH^{*})(1-\hat{m}^2_{\phi})\right]
+|H|^2\frac{\hat{m}^2_l}{\hat{m}^2_{\phi}}\hat{s}\lambda
\label{ims2} \ea

The normalized forward-backward asymmetries (FBA) is defined as
\ba {\cal A}_{FB}(\hat{s})=\int
d\hat{s}~\frac{\int^{+1}_{-1}dcos\theta\frac{d^2Br}{d\hat{s}dcos\theta}{\rm
Sign}(cos\theta)}
{\int^{+1}_{-1}dcos\theta\frac{d^2Br}{d\hat{s}dcos\theta}}.\ea
According to this definition, the explicit expressions of FBA for
the exclusive decays is:
\ba\label{fbaks} \frac{d{\cal
A}_{FB}}{d\hat{s}}&=&\frac{1}{D^{\phi}}\hat{u}(\hat{s})\Bigg \{
\hat{s}[Re(BE^{*}) +Re(AF^{*})] \non
& & +\frac{\hat{m}_l}{\hat{m}_{\phi}}[Re({\cal S}_2B^{*})
(1-\hat{s}-\hat{m}^2_{\phi})-Re({\cal S}_2 C^{*})\lambda]\Bigg \}.
\ea

Now we are ready to present the analytical expressions of lepton
polarization by defining: \be
\frac{d\Gamma(\hat{n})}{d\hat{s}}=\frac{1}{2}\big
(\frac{d\Gamma}{d\hat{s}}\big )_0[1
+(P_L\hat{e}_L+P_N\hat{e}_N+P_T\hat{e}_T)\cdot\hat{n}] \ee where
the subscript $"0"$ corresponds to the unpolarized decay case,
$P_L$ and $P_T$ are the longitudinal and transverse polarization
asymmetries in the decay plane respectively, and $P_N$ is the
normal polarization asymmetry in the direction perpendicular to
the decay plane.

The lepton polarization asymmetry $P_i$ can be derived by \be
P_i(\hat{s})=\frac{d\Gamma(\hat{n}=\hat{e}_i)/d\hat{s}-
d\Gamma(\hat{n}=-\hat{e}_i)/d\hat{s}}{d\Gamma(\hat{n}=\hat{e}_i)/d\hat{s}+
d\Gamma(\hat{n}=-\hat{e}_i)/d\hat{s}}\;  \ee
the results are
\ba
P_L &=&\frac{1}{D^{\phi}}{\cal D}\Bigg\{\frac{2\hat{s}\lambda}{3}
Re(AE^{*})+\frac{(\lambda+12\hat{s}\hat{m}^2_{\phi}
)}{3\hat{m}^2_{\phi}}Re(BF^{*})\Bigg.  \non
&&\Bigg.-\frac{\lambda(1-\hat{m}^2_{\phi}-\hat{s})}{3\hat{m}^2_{\phi}}Re(BG^{*}+CF^{*})
+\frac{\lambda^2}{3\hat{m}_{\phi}}Re(CG^{*})\Bigg.\non
&&\Bigg.+\frac{2\hat{m}_l\lambda}{\hat{m}_{\phi}}[Re(F{\cal
S}^{*}_2)-
\hat{s}Re(H{\cal S}^{*}_2)-(1-\hat{m}^2_{\phi})Re(G{\cal S}^{*}_2)]\Bigg\},\label{plks}\\
P_N &=&\frac{1}{D^{\phi}}\frac{-\pi\sqrt{\hat{s}}\hat{u}(\hat{s})}{4\hat{m}_{\phi}}\Bigg\{
\frac{\hat{m}_l}{\hat{m}_{\phi}}\left[Im(FG^{*})(1+3\hat{m}^2_{\phi}-\hat{s})
\right.\Bigg. \non
&&\Bigg.\left.+Im(FH^{*})(1-\hat{m}^2_{\phi}-\hat{s})-Im(GH^{*})\lambda
\right]\Bigg. \non
&&\Bigg.+2\hat{m}_{\phi}\hat{m}_l[Im(BE^{*})+Im(AF^{*})]\Bigg.\non
&&\Bigg.-(1-\hat{m}^2_{\phi}-\hat{s})Im(B{\cal S}^{*}_2)+\lambda
Im(C{\cal S}_2^{*})\Bigg\},\label{pnks}\\
P_T &=&\frac{1}{D^{\phi}}\frac{\pi\sqrt{\lambda}\hat{m}_l}{4\sqrt{\hat{s}}}
\Bigg\{ 4\hat{s}Re(AB^{*})\Bigg.\non
&&\Bigg.+\frac{(1-\hat{m}^2_{\phi}-\hat{s})}{\hat{m}^2_{\phi}}\left[
-Re(BF^{*})+(1-\hat{m}^2_{\phi})Re(BG^{*})+\hat{s}Re(BH^{*})\right]\Bigg.\non
&&\Bigg.+\frac{\lambda}{\hat{m}^2_{\phi}}[Re(CF^{*})-(1-\hat{m}^2_{\phi})Re(CG^{*})
-\hat{s}Re(CH^{*})]\Bigg.\non
&&\Bigg.+\frac{(\hat{s}-4\hat{m}^2_l)}{\hat{m}_{\phi}\hat{m}_l}
[(1-\hat{m}^2_{\phi}-\hat{s})Re(F{\cal S}^{*}_2)-\lambda Re(G{\cal
S}^{*}_2)]\Bigg\}. \label{ptks}
\ea
where ${\cal D}=\sqrt{1-4\frac{\hat{m}^2_{l}}{\hat{s}}}$,
$D^{\phi}$ are given in Eq.~(\ref{ims2}).

\section{Numerical result}
In the numerical calculations, we fix the SM parameters as follows~\cite{UTfitCKM,PDG10,PMass}.
\begin{gather}
A=0.8095,\
\lambda=0.22545, \ \overline{\rho}=0.132\pm0.02,\
\overline{\eta}=0.367\pm0.013. \nonumber \\
\ m_c = 1.4 \mbox{ GeV}, \ m_{b} = 4.8 \mbox{ GeV}, \ m_{t}=172.4\mbox{ GeV},\
 \nonumber \\
 m_{\mu} = 0.1057 \mbox{ GeV}, \  m_{\tau} = 1.7769 \mbox{ GeV} \ m_{W}=80.4\mbox{ GeV},\ \nonumber\\
  m_{Z}=91.18\mbox{GeV},\ m_{B_s} = 5.36 \mbox{ GeV},\  m_{\phi} = 1.02 \mbox{ GeV}, \ \nonumber\\
\alpha_{em}=\frac{1}{137},\ \alpha_s(m_Z)=0.118, \ sin^2 \theta_W
=0.23,\ \tau_{B_s}=1.46\times10^{-12}\mbox{s}. \label{parameter}
\end{gather}

In the TC2 model, the new physics contributions depend on new parameters which have
been constrained by theory arguments and by experimental results. $\varepsilon$ denotes the portion
of the top quark mass generated by the extended technicolor. The experimental constraints on $\varepsilon$
from the data of radiative decay $b \to s \gamma$ are weak~\cite{bsgam}. However, from the theoretical point
of view, $\varepsilon$ is favored in the range of $[0.03,0.1]$~\cite{Hill}.

On the theoretical side, Ref.~\cite{Hill} estimated that the mass of top-pions should be a few hundred GeV using
quark loop approximation, and Refs.~\cite{Hill,tc204} evaluated the mass of top-Higgs to be about $2m_t$.
On the experimental side, the neutral top-pion and the top-Higgs are weakly restricted. Meanwhile, the
mass of the charged top-pion have been strongly constrained. For example, the absence of $t \to \pi_t^+ b$
indicates that $m_{\pi_t^+}>165{\rm GeV}$~\cite{Balaji}, and the analysis of $R_b$ reveals that
$m_{\pi_t^+}>220{\rm GeV}$~\cite{Burdman,Hill95}.

As for the bounds on the mass of $Z'$, precision electroweak data show that $m_{Z'}$ must be larger
than $1{\rm TeV}$~\cite{Chivukula}. The vacuum tilting, the confinement
from Z-pole physics, and U(1) triviality need $K_1 \leq 1$~\cite{Popovic}.
When considering experimentally much better measured modes such as $B \to \mu^+ \mu^-$ and $B \to K l^+ l^-$,
we can easily obtain the constraints on the free parameters $m_{Z'}$ and $K_1$. For example,
for $K_1 =0.4$, we must have $1290{\rm GeV} < m_{Z'} < {\rm 1787GeV}$~\cite{chongxingyue99}.

\begin{figure}[t]
\begin{center}
\vspace{-2.5cm}
\epsfxsize=20cm \centerline{\epsffile{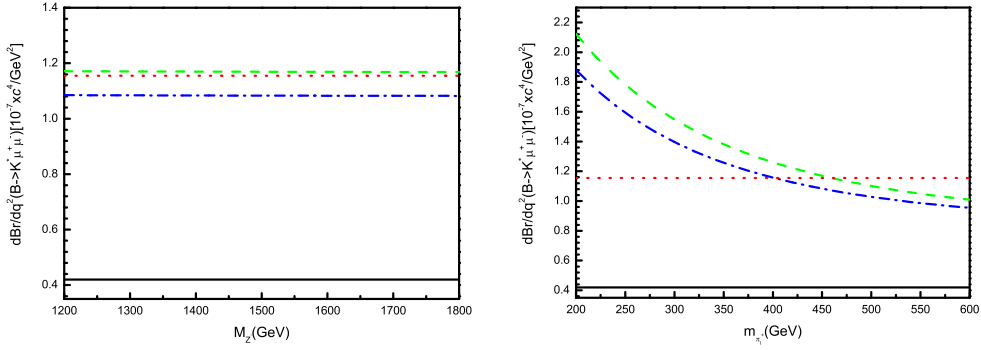}}
\vspace{-8.5cm }
\caption{\label{Br_ksmumzpmpi}\small Plots of differential branching fractions
$dBr/dq^2(B\to K^* \mu^+\mu^-)$
at low $q^2$ in function of the mass $M_{Z'}$ (left panel) and of the
mass $m_{\pi_t^+}$ (right panel). }
\vspace{-1cm}
\end{center}
\end{figure}

The differential branching fraction of exclusive decay $B \to K^{*} \mu^+ \mu^-$ has been already measured by BaBar, Belle, CDF and LHCb.
The latest LHCb result which corresponds to an integrated luminosity of 1 ${\rm fb}^{-1}$ in the low $q^2$ region is~\cite{LHCb-ksmumu}
\be
\left[\frac{dBr}{dq^2}(B \to K^{*} \mu^+ \mu^-)\right]_{[1,6]} = (0.42 \pm 0.04 \pm 0.04 )\times 10^{-7} c^4/{\rm GeV^2}.
\label{brsm}
\ee
With the above precise measurement, we give the plots of differential branching fraction $dBr/dq^2(B\to K^* \mu^+\mu^-)$
at low $q^2$ in function of the mass $M_{Z'}$ (left panel) and of the mass $m_{\pi_t^{+}}$ (right panel)
in Fig.~\ref{Br_ksmumzpmpi}. The solid lines denote the LHCb central value, while the dotted lines show
the $3\sigma$ bound including the experimental errors with theoretical ones
given in Table 5 of Ref.~\cite{ksmumu-theory} (added in quadrature). The dashed and
dash-dotted curve corresponds to the TC2 prediction for $\varepsilon=0.04$ and
$\varepsilon=0.08$, respectively.
It is easy to see that the whole parameter space of $M_{Z'}$ is excluded
for $\varepsilon=0.04$, but allowed for
$\varepsilon=0.08$ by this differential branching fraction.
The mass of top-pion $\pi_t^{+}$ below 450 GeV for $\varepsilon=0.04$
and below 400 GeV for $\varepsilon=0.08$ are also excluded by the LHCb data.

After taking into account the new constraints from $B\to K^* \mu^+\mu^-$ decay,
we will make numerical calculations by using the input parameters in the following ranges:
\ba
m_{\pi_t^+}&=&(350-600){\rm GeV},\quad
m_{\pi_t^0}=m_{h_t^0}=(200-500){\rm GeV}, \quad
m_{Z'}=(1200-1800){\rm GeV}, \label{parameter2} \non
\varepsilon&=&(0.06-0.1),\quad K_1=(0.3-1), \quad F_\pi=50 {\rm GeV}.
\ea

Using the above input parameters, we will calculate the physics
observables as defined in previous sections
and study the sensitivity to the new physics corrections appeared
in the TC2 model. The invariant mass spectra and branching ratios are almost the
same for electron and muon modes because the mass of electron and
muon are small. Meanwhile, the electron polarization is very difficult to
measure, so we only consider $B_s \to \phi \mu^+\mu^-, \phi \tau^+\tau^-$ decays in this work.

Using Eq.~(\ref{ims}) and the input parameters as given above, it is easy to calculate the branching ratio
$Br(B_s \to \phi \mu^+\mu^-, \phi \tau^+\tau^-)$. In the SM, the numerical results are
\ba
Br(B_s \to \phi \mu^+\mu^-) &=& 1.54^{+0.28}_{-0.25}\times 10^{-6}, \non
Br(B_s \to \phi \tau^+\tau^-) &=& 1.65^{+0.30}_{-0.28}\times 10^{-7}.
\label{brsm2}
\ea
where the error corresponds to the uncertainty of input parameters of form factors.

In the TC2 model, both the new penguin and tree level diagrams contribute through constructive interference with
their SM counterparts and consequently provide large enhancements with respect to the SM predictions.
For the typical values of $F_{\pi}=50GeV$, $\varepsilon=0.08$, $K_1=0.4$, $m_{\pi^{+}_t}=450GeV$,
$m_{\pi^0_t}=m_{h^0_t}=300GeV$ and $M_{Z'}=1500GeV$, one has
\ba
Br(B_s \to \phi \mu^+\mu^-) &=& \left \{
\begin{array}{ll}
1.55\times 10^{-6} & {\rm only}\ \ Z'\ \ {\rm considered}, \\
3.07\times 10^{-6} & {\rm only}\ \ \pi_t^{+} \ \ {\rm considered}, \\
3.76\times 10^{-6} & {\rm both}\ \ Z' \ {\rm and}\ \ \pi_t^{+}
\ \  {\rm considered}.
\end{array} \right.\\
Br(B_s \to \phi \tau^+\tau^-) &=& \left \{
\begin{array}{ll}
1.80\times 10^{-7} & {\rm only}\ \ Z'\ \ {\rm considered}, \\
2.92\times 10^{-7} & {\rm only}\ \ \pi_t^{+} \ \ {\rm considered}, \\
3.14\times 10^{-7} & {\rm both}\ \ Z' \ {\rm and}\ \ \pi_t^{+}
\ \  {\rm considered}.
\end{array} \right.
\label{brtc}
\ea
\begin{figure}[t]
\begin{center}
\vspace{-2.5cm}
\epsfxsize=20cm \centerline{\epsffile{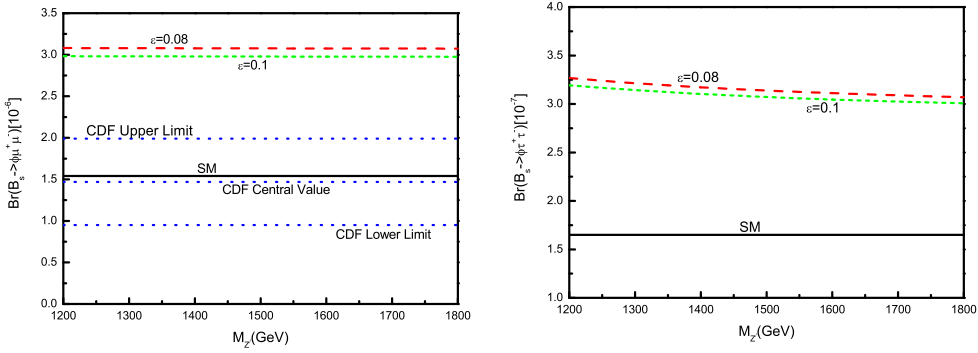}} \vspace{-8.5cm}
\centerline{\parbox{16cm}{\caption{\label{Br_mutaumzp}\small Plots of branching ratios of $Br(B_s\to \phi \mu^+\mu^-, \phi \tau^+\tau^-)$
decays versus $M_{Z'}$ in the SM and TC2 model. The dashed, short-dashed lines and solid curve correspond to the TC2 and SM results, respectively. The dotted lines denote the CDF data with $1\sigma$ error: $Br(B_s\to \phi \mu^+\mu^-)=(1.47\pm 0.52) \times 10^{-6}$.}}}
\vspace{-1cm}
\end{center}
\end{figure}

In Fig.~\ref{Br_mutaumzp}, we show the branching
ratios of decays $B_s\to \phi \mu^+\mu^-, \phi \tau^+\tau^-$ as a function
of the mass parameter $M_{Z'}$ in the SM and TC2 model. The solid line refers to the SM prediction,
while the dashed, short-dashed curves correspond to theoretical prediction with the inclusion of
the new physics effects of the TC2 model for $\varepsilon=0.08$ and $\varepsilon=0.1$, respectively.
The dotted lines denote the CDF data with $1\sigma$ error. From this figure,
we can see that the new physics enhancements can be significant in size.
For $B_s\to \phi \mu^+\mu^-$ decay mode, the values of branching ratio basically remain unchanged within the range of
$M_{Z'}=1200\sim 1800 GeV$. The theoretical predictions of $Br(B_s \to \phi \tau^+\tau^-)$ have some sensitivity
to the parameter $M_{Z'}$ because the nonuniversal gauge boson $Z'$ has large couplings to the third generation
fermion with respect to the first two generations.
\begin{figure}[t]
\begin{center}
\vspace{-2cm}
\epsfxsize=20cm \centerline{\epsffile{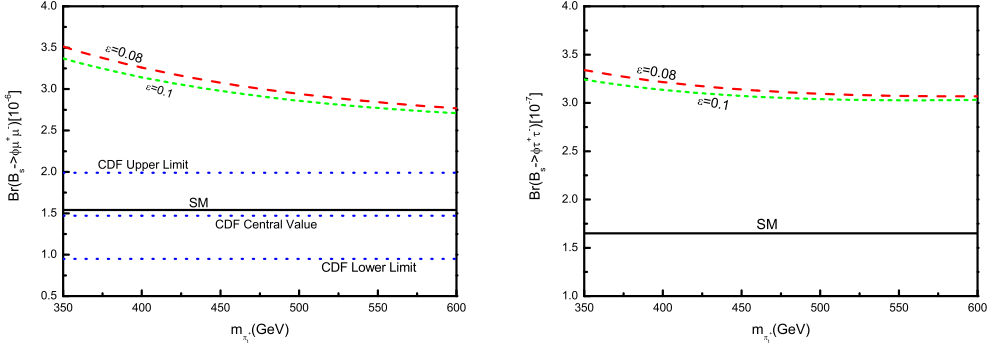}} \vspace{-8.7cm}
\centerline{\parbox{16cm}{\caption{\label{Br_mutaumpi}\small Plots of branching ratios of $Br(B_s\to \phi \mu^+\mu^-, \phi \tau^+\tau^-)$
decays versus $m_{\pi_t^{+}}$ in the SM and TC2 model. The dashed, short-dashed lines and solid curve represent the TC2 and SM results, respectively.}}}
\end{center}
\end{figure}

In Fig.~\ref{Br_mutaumpi}, we show the branching ratios of decays
$B_s\to \phi \mu^+\mu^-, \phi \tau^+\tau^-$ as a function of the mass parameter $m_{\pi_t^{+}}$ in the SM and TC2 model.
One can see from Fig.~\ref{Br_mutaumpi} that the new physics enhancements to the two studied decays
are still large in size when the parameter $m_{\pi_t^{+}}$ varies. The branching ratios
$Br(B_s\to \phi \mu^+\mu^-, \phi \tau^+\tau^-)$ are not very sensitive to the variations of the input parameter $\varepsilon$.
For $\varepsilon=0.08$ and $\varepsilon=0.1$,
the enhancement to the $Br(B_s\to \phi \mu^+\mu^-, \phi \tau^+\tau^-)$ can reach a factor of $\sim 2$.
The uncertainty of the data is still large. Further improvement of the data will be very helpful to test
or constrain the parameter $m_{\pi_t^{+}}$ in the TC2 model from these decays.
\begin{figure}[t]
\begin{center}
\vspace{-2.5cm}
\epsfxsize=20cm \centerline{\epsffile{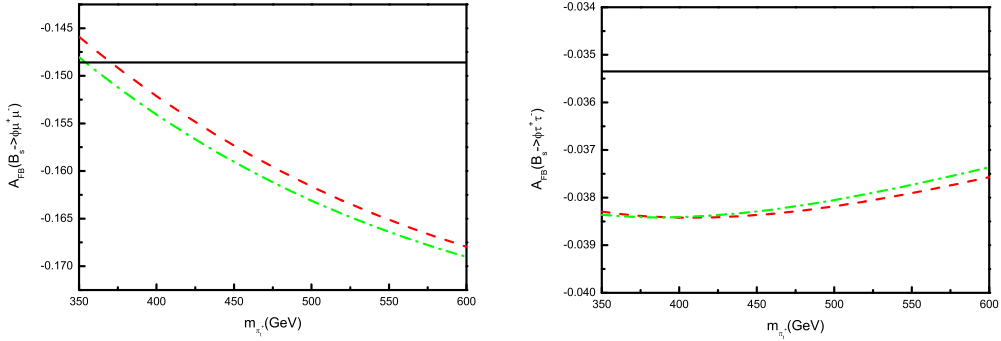}} \vspace{-7.8cm}
\centerline{\parbox{16cm}{\caption{\label{AFB_mutaumpi}\small Plots of forward-backward asymmetries of $A_{FB}(B_s\to \phi \mu^+\mu^-, \phi \tau^+\tau^-)$ decays versus $m_{\pi_t^{+}}$ in the SM and TC2 model. The dashed, dash-dotted lines and solid curve stand for the TC2 and SM results, respectively.}}}
\end{center}
\end{figure}

In Fig.~\ref{AFB_mutaumpi}, we show the forward-backward asymmetries of decays
$B_s\to \phi \mu^+\mu^-, \phi \tau^+\tau^-$ as a function of the mass parameter $m_{\pi_t^{+}}$ in the SM and TC2 model.
The solid line denotes the SM prediction,
while the dashed, dash-dotted curves correspond to the theoretical prediction of TC2 model for
$\varepsilon=0.08$ and $\varepsilon=0.1$, respectively. For $B_s\to \phi \mu^+\mu^-$ decay, the
theoretical prediction of the forward-backward asymmetry in the SM is: $A_{FB}(B_s\to \phi \mu^+\mu^-)=-0.149 \pm 0.001$.
In the TC2 model, when the $\pi_t^{+}$ mass is in the range of $350GeV \sim 600GeV$, the value of $A_{FB}(B_s\to \phi \mu^+\mu^-)$
is in the range of $-0.168 \sim -0.146$ for $\varepsilon=0.08$.
For $B_s\to \phi \tau^+\tau^-$ decay, the forward-backward asymmetry amounts to $-0.038 \sim -0.037$
for $\varepsilon=0.08$, which is comparable with the SM result of $-0.035\pm 0.0001$.

\begin{figure}[t]
\begin{center}
\vspace{-2.5cm}
\epsfxsize=20cm \centerline{\epsffile{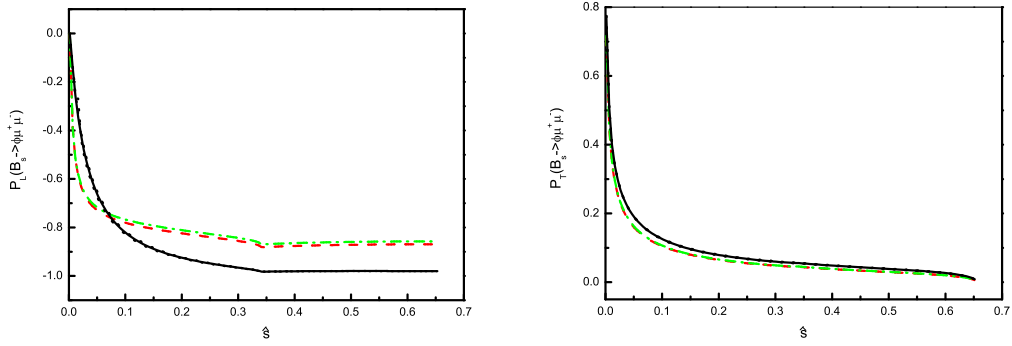}} \vspace{-8.2cm}
\centerline{\parbox{16cm}{\caption{\label{PLPT_mus}\small Plots of
$P_L$ and $P_T$ for decay $B_s\to \phi \mu^+\mu^-$ in the SM and TC2 model.
The dashed, dash-dotted lines and solid curve display the central values of the TC2 and SM predictions, respectively.
The two dotted lines show the uncertainties of form factors induced by F(0) in the SM.}}}
\end{center}
\end{figure}

\begin{figure}[t]
\begin{center}
\vspace{-2.0cm}
\epsfxsize=20cm \centerline{\epsffile{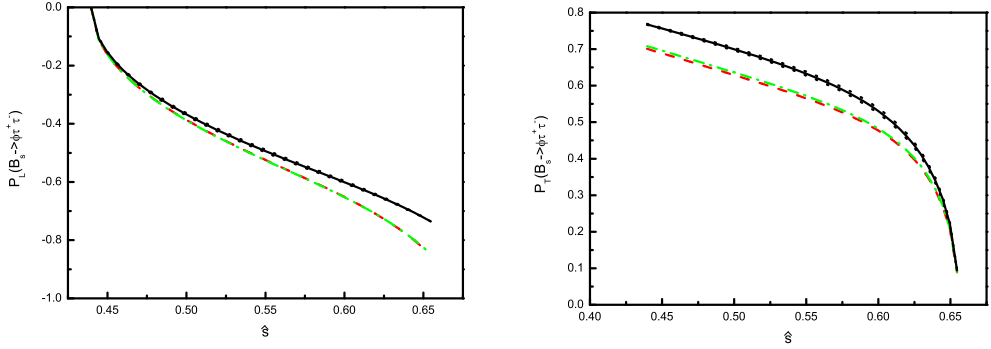}} \vspace{-8.5cm}
\centerline{\parbox{16cm}{\caption{\label{PLPT_taus}\small Plots of
$P_L$ and $P_T$ for decay $B_s\to \phi \tau^+\tau^-$ in the SM and TC2 model.
Other captions are same as Fig.~\ref{PLPT_mus}.}}}
\end{center}
\end{figure}

In Figs.~\ref{PLPT_mus} and \ref{PLPT_taus}, we present the longitudinal and transverse
polarization for decays $B_s\to \phi \mu^+\mu^-$ and $B_s\to \phi \tau^+\tau^-$.
The solid line is the SM prediction, while the dashed, dash-dotted curves are the theoretical prediction of TC2 model for
$\varepsilon=0.08$ and $\varepsilon=0.1$, respectively. From these figures, it easy to see that the variations of the
input parameter $\varepsilon$ can only provide a few percent change of the lepton polarization for
$B_s\to \phi \mu^+\mu^-, \phi \tau^+\tau^-$ decays. For the decay $B_s\to \phi \mu^+\mu^-$, the $P_L$ is
suppressed by about $110\%$ at most at the small momentum transfer, while at $\hat{s}>0.07$, it will become larger than that
of the SM, and the new physics contribution in TC2 model provide an enhancement of $\sim 12\%$. For the $P_T$ part,
the new physics contribution result in a $(8\sim18)\%$ decrease. As for $B_s\to \phi \tau^+\tau^-$, the deviation from the
SM prediction appears when $\hat{s}>0.5$ for $P_L$. $P_T$ is decreased with respect to the SM prediction by about $10\%$
in all the di-lepton invariant mass range. Thus, the measurement of $P_L$ for $B_s\to \phi \mu^+\mu^-$ and $P_T$ for
$B_s\to \phi \tau^+\tau^-$ will distinguish between the SM and the TC2 model.

\section{Summary}
In this paper, we carried out a study of the new physics contributions to the branching ratios, forward-backward asymmetries and
lepton polarization for the decays $B_s\to \phi \mu^+\mu^-, \phi \tau^+\tau^-$ in the TC2 model by using form factors calculated
within the light-cone QCD sum rule approach.

In Section~II, a brief review about the topcolor-assisted
technicolor model was given. In Section~III, we presented the theoretical framework for $B_s\to \phi l^+ l^-$
decays within the TC2 model, then give the definitions and the derivations of the form
factors in the decays $B_s \to \phi l^+ l^-$ using the updated form factors within the light-cone
QCD sum rule. In Section~IV, we introduced the basic formula for experimental observables.
In Section~V, we calculated the branching ratio, forward-backward asymmetry, and lepton
polarization of $B_s \to \phi l^+ l^-$ and made phenomenological analysis for these decays in the SM and
the TC2 model. From the numerical results, we found the following features about the new physics effects:

\begin{itemize}
\item The branching ratios of $\bar{B}_s\to \phi \mu^+\mu^-, \phi \tau^+\tau^-$ decays are essentially unaffected
by the $Z^{\prime}$ contributions, while charged top-pions interaction can lead to striking effects in these decay
distributions. For $\varepsilon=0.08$ and $\varepsilon=0.1$, the enhancement can reach a factor of $\sim 2$.

\item For the forward-backward asymmetry of the decay $B_s\to \phi \mu^+\mu^-$, the
NP enhancement is in the range $-13\%$ to $3\%$. For $B_s\to \phi \tau^+\tau^-$ decay, the
NP effects is about $-9\%$ to $-6\%$ compared to the SM predictions.

\item For the lepton polarization, $P_L(B_s\to \phi \mu^+\mu^-)$ is
increased by about $12\%$. However, $P_T(B_s\to \phi \mu^+\mu^-)$
is decreased by $(8\sim18)\%$. As for $B_s\to \phi \tau^+\tau^-$, the deviation from the
SM prediction appears when $\hat{s}>0.5$ for $P_L$. In the $P_T$ part, the SM prediction
will be decreased by about $10\%$.

\end{itemize}

An improved measurement of $Br(\bar{B}_s\to \phi \mu^+\mu^-)$ and first measurements of
the longitudinal polarization asymmetry, $P_L$, in $B_s\to \phi \mu^+\mu^-$ and of
the transverse polarization asymmetry, $P_T$, in
$B_s\to \phi \tau^+\tau^-$ at LHCb and super-flavor factories (BellII and the proposed Super-B )
will allow to distinguish between the SM and the TC2 model.

\section*{Acknowledgments}
The authors would like to thank Prof. Zhen-jun Xiao for helpful
comments and suggestions on the manuscript.
The work is supported by the National Science Foundation
under contract No.~10947020, and Natural Science Foundation of
Henan Province under Grant No.~112300410188.

\begin{appendix}

\section{Relevant functions in the TC2 model}\label{app:aa}
In this Appendix, we give the explicit expressions of functions that related to the
rare B decays studied here in the framework of the TC2 model.
\ba
K_{ab}(x)&=&-\frac{2g^2c_w^2I_1(x)}{3g_2^2(v_d+a_d)},\\
K_{c}(x)&=&\frac{2f^2c_w^2}{g_2^2}\left[\frac{2I_2(x)}{3(v_u+a_u)}+\frac{I_3(x)}{6(v_u-a_u)}\right],\\
K_{d}(x)&=&\frac{2f^2c_w^2}{g_2^2}\left[\frac{2I_4(x)}{3(v_u+a_u)}+\frac{I_5(x)}{6(v_u-a_u)}\right],\\
C(x)&=&\frac{I_1(x)}{-[0.5(Q-1)s_w^2+0.25]}.
\ea
Here $g=\sqrt{4\pi K_1}$, $s_w=\sin \theta _{w}$, $a_{u,d}=I_3$,
$v_{u,d}=I_3-2Q_{u,d}s_w^2$,
and $u, d$ stand for the up and down type quarks, respectively.
\ba
I_1(x)&=&-(0.5(Q-1)s_w^2+0.25)(x^2 ln(x)/(x-1)^2-x/(x-1)
-x(0.5(-0.5772\nonumber\\
& &+ln(4\pi)-ln(M_W^2))
+0.75-0.5(x^2ln(x)/(x-1)^2-1/(x-1)))),\\
I_2(x)&=&(0.5 Q s_w^2-0.25)(x^2ln(x)/(x-1)^2-2x
ln(x)/(x-1)^2+x/(x-1)),\\
I_3(x)&=&-Q s_w^2(x/(x-1)-x ln(x)/(x-1)^2),\\
I_4(x)&=&0.25(4 s_w^2/3-1)(x^2ln(x)/(x-1)^2-x-x/(x-1)),\\
I_5(x)&=&-0.25Q s_w^2x(-0.5772+ln(4\pi)-ln(M_W^2)+1-x
ln(x)/(x-1))\nonumber\\
& &-s_w^2/6(x^2ln(x)/(x-1)^2-x-x/(x-1)).
\ea

\end{appendix}

\end{document}